\documentclass[a4paper]{jpconf}
\usepackage{graphicx}
\begin{document}
\title{Large Angle Hadron Correlations from Medium-Induced 
       Gluon Radiation}

\author{Ivan Vitev}

\address{Los Alamos National Laboratory, Theory Division and Physics Division,
Mail Stop H846,\\  Los Alamos, NM 87544}

\ead{ivitev@lanl.gov}

\begin{abstract}
Final state medium-induced  gluon radiation in ultradense 
nuclear matter is shown to favor large angle emission 
when compared to vacuum bremsstrahlung due to the suppression of 
collinear gluons. In a simple model of momentum transfers from the 
medium to the jet+gluon system the phase space distribution
of non-Abelian bremsstrahlung  is examined  to first order in opacity.  
Perturbative expression for the contribution of the hadronic fragments 
from medium-induced  gluons to the back-to-back particle correlations 
is derived.  It is found that in the limit of large jet energy loss  
gluon radiation  determines the yield and angular distribution 
of $|\Delta \varphi | \geq \frac{\pi}{2}$ di-hadrons to high 
transverse momenta  $p_{T_2}$ of the associated particles. 
Clear transition from enhancement to suppression of the away-side 
di-hadron correlations is established at moderate $p_{T_2}$ and its 
experimentally accessible features are predicted versus the trigger 
particle momentum $p_{T_1}$. The measured 
away-side correlation function is shown to be sensitive to 
the broad rapidity distribution of di-jets and their vacuum and 
medium-induced acoplanarity.
\end{abstract}

\section{Introduction}

The discovery of jet quenching~\cite{Levai:2001dc} -- 
the suppression of large transverse momentum  hadron  
production in nuclear collisions relative to the expectation from  
p+p reactions scaled  by the number of elementary nucleon-nucleon 
interactions -- is arguably the most exciting new result 
from the Au+Au experimental program at the Relativistic 
Heavy Ion Collider (RHIC). The phenomenon of jet quenching 
has been established via the attenuation of the single 
inclusive particle spectra~\cite{Adler:2003qi} 
and the suppression of the back-to-back di-hadron
correlations~\cite{Adler:2002tq,unknown:2005ph}. It has been 
interpreted as critical evidence for large parton energy 
loss~\cite{Gyulassy:2000gk,Wang:2003mm,Adams:2003im} 
in ultradense quark-gluon plasma (QGP) -- 
the deconfined state of matter predicted by quantum 
chromodynamics (QCD).

So far, the observable effects of the medium-induced gluon 
radiation {\em itself} were considered to be modest. For tagged 
jets, assuming thermalization of the lost energy calculated 
in~\cite{Gyulassy:2000fs}, transport models predicted an 
increase in the multiplicity of associated particles limited to 
$p_{T_2} \leq 500$~MeV~\cite{Pal:2003zf}. Using 
the angular gluon distribution from~\cite{Baier:1999ds}, jet cone 
broadening was found to be $ < 10 \% $ and challenging to detect 
experimentally even at the CERN  Large Hadron Collider 
(LHC)~\cite{Salgado:2003rv}.

In these proceedings we follow Ref.~\cite{Vitev:2005yg} and 
demonstrate that a mechanism, based on the  destructive interference 
of color currents from hard and soft parton scattering, can 
ensure broad in angle and frequency final state 
medium-induced emission spectrum in QCD. For large energy 
loss, perturbative fragmentation of the radiative gluons is found 
to give a dominant contribution to the yield of away-side di-hadrons 
and to significantly alter their correlations at transverse momenta 
much higher than naively anticipated.  We note that the 
medium-modified jet topology can be distorted in two particle 
back-to-back measurements by the broad di-jet rapidity distribution
and the vacuum and medium-induced acoplanarity. If present, it should
be  more cleanly detectable on an event-by-event basis and 
through jet energy flow measurements.

\section{Angular distribution of the medium-induced 
non-Abelian bremsstrahlung}

\begin{figure}[!t]
\begin{center}
\includegraphics[width=3.in,angle=0]{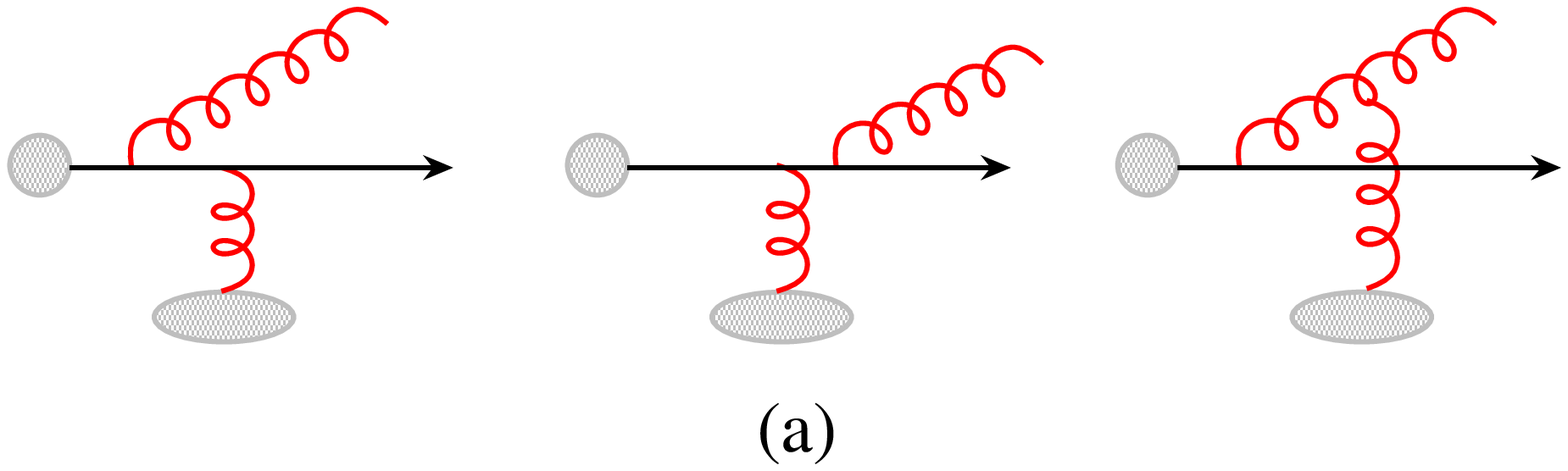}  \hspace*{.5cm}
\includegraphics[width=3.in,angle=0]{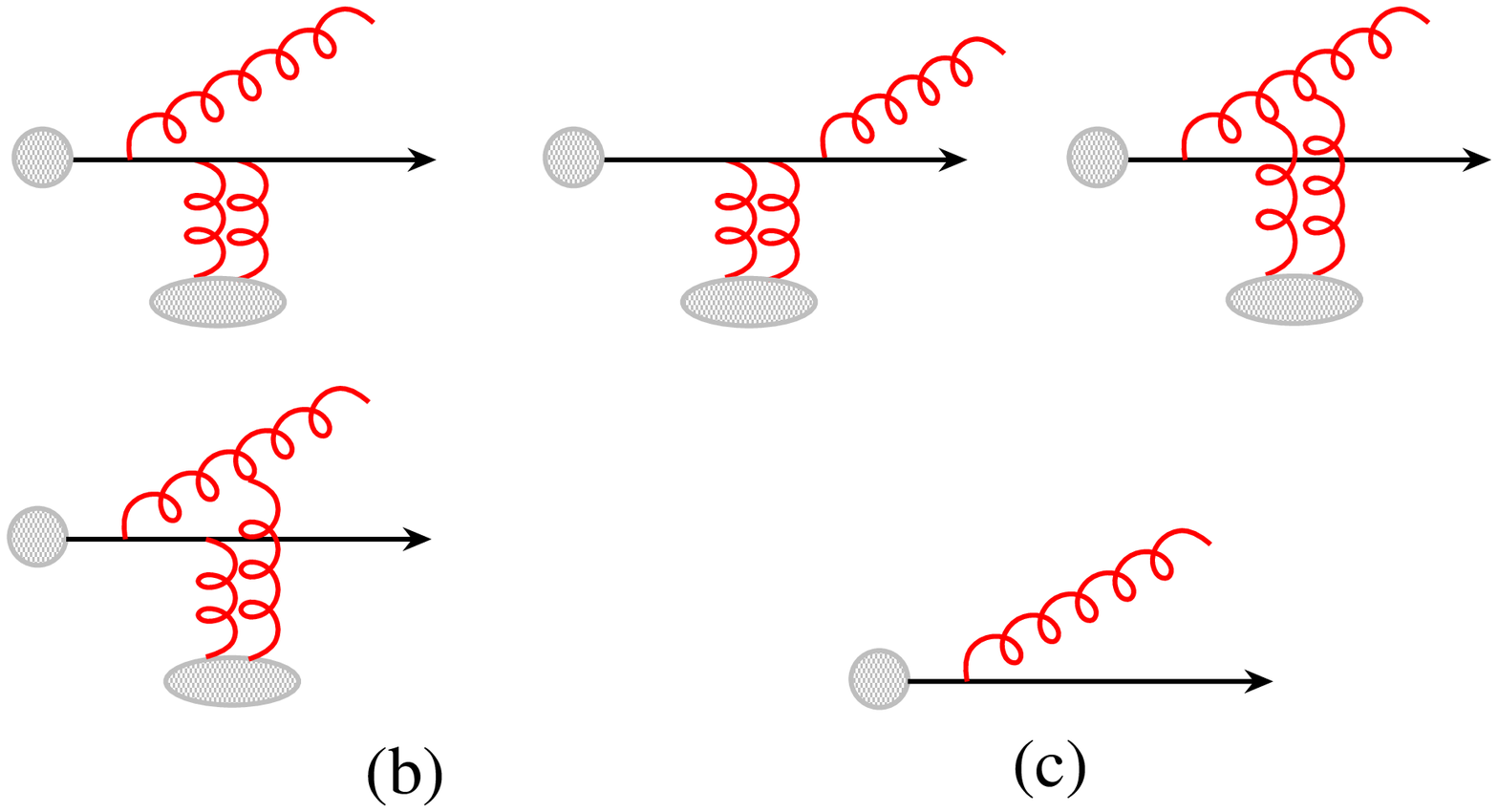}
\end{center}
\caption{ (a) Direct diagrams that contribute to first order in
opacity $\bar{n} = L/\lambda$. (b) Double Born diagrams that
interfere with the vacuum  bremsstrahlung (c) to the same 
order in $\bar{n}$.}
\label{diags}
\end{figure}

\begin{figure}[!t]
\begin{center}
\includegraphics[width=4.5in,height=3.3in,angle=0]{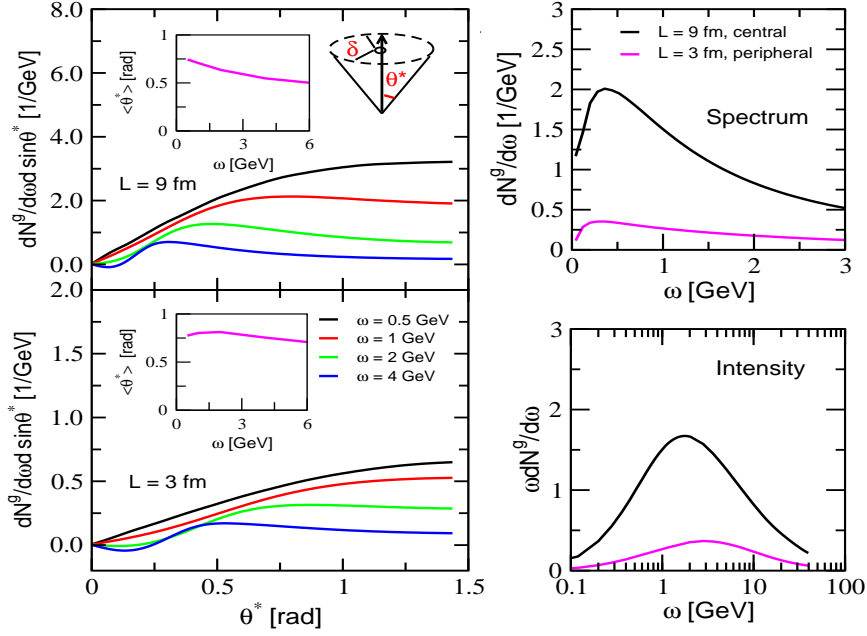}
\end{center}
\caption{ The angular  distribution of medium-induced 
bremsstrahlung  of  $E = 6$~GeV gluon jet  for fixed values of 
the radiative gluon energy  $\omega = 0.5, 1, 2, 4$~GeV. 
Top and bottom panels 
represent (1+1)D Bjorken expanding medium of transverse size 
$L=9$~fm and $L=3$~fm, respectively.
Inserts show  $\langle \theta^* \rangle$ versus $\omega$. 
Right panels illustrate the gluon spectrum
and radiation intensity of $E = 40$~GeV gluon
jet.}
\label{hole}
\end{figure}

We first recall that for hard perturbative scattering 
the radiative spectrum of  real gluon emission for small and  
moderate frequencies $\omega$ 
is given by~\cite{Gyulassy:2000fs}  
\begin{equation} 
|{\cal M}_c|^2 \propto 
\frac{dN^g_{\rm vac}}{d\omega d \sin \theta^* d \delta } \approx  
\frac{ C_R \alpha_s} { \pi^2} 
\frac{ 1 }{ \omega \sin  \theta^* }   \;. 
\label{divsmall} 
\end{equation}
The distribution in Eq.~(\ref{divsmall}) corresponds to the
diagram in Fig.~\ref{diags}c where the blob represents the large
virtual momentum exchange.
In Eq.~(\ref{divsmall}) $\theta^* = \arcsin(k_\perp/\omega)$ is 
the angle relative to the jet axis,  $\delta$ is the azimuthal 
cone angle (both illustrated in Fig.~\ref{hole}),   
$\alpha_s$ is the strong coupling constant and 
$C_R = 3 \, (4/3)$ for gluon (quark) jets, respectively. 
Virtual gluon corrections remove the $\omega \rightarrow 0$ 
infrared singularity in the cross sections 
in accord with the  Kinoshita-Lee-Nauenberg 
theorem~\cite{Kinoshita:1962ur} but the collinear   
$\theta^* \rightarrow 0$  divergence has to be regulated or subtracted 
in the parton distribution functions (PDFs) and the fragmentation 
functions (FFs). 

In contrast, the final state medium-induced bremsstrahlung spectrum 
is both collinear and infrared safe.
The full solution for the medium induced gluon radiation
off jets produced in a hard collisions at early 
times $\tau_{jet} \simeq 1/E$ inside a  nuclear  
medium  of length $L$  can be obtained to all orders  
in the correlations  between the  multiple  scattering  centers  
via the reaction operator approach~\cite{Gyulassy:2000fs}
\begin{eqnarray}
 \frac{dN^g_{\rm med}}{d \omega \, d \sin \theta^* d \delta }  
&=&
\sum\limits_{n=1}^\infty  
\frac{dN^{g\, (n)}_{\rm med}}{d \omega \, d \sin \theta^* d \delta }  
\;  =  \; \omega \sin \theta^*  
\sum\limits_{n=1}^{\infty}  \frac{2 C_R \alpha_s}{\pi^2} 
 \; \prod_{i=1}^n \;\int_0^{L-\sum_{a=1}^{i-1} \Delta z_a } 
 \frac{d \Delta z_i }{\lambda_g(i)} 
 \,  \nonumber \\[1.ex] 
&& \times \int   d^2{\bf q}_{i} \, 
\left[  \sigma_{el}^{-1}(i) \frac{d \sigma_{el}(i)}{d^2 {\bf q}_i}
  - \delta^2({\bf q}_{i}) \right]  \,  
\left( - \,{\bf C}_{(1, \cdots ,n)} \cdot 
\sum_{m=1}^n {\bf B}_{(m+1, \cdots ,n)(m, \cdots, n)}  \right.
\nonumber \\[1.ex] 
&\;& \times 
\left.
\left[ \cos \left (
\, \sum_{k=2}^m \omega_{(k,\cdots,n)} \Delta z_k \right)
-   \cos \left (\, \sum_{k=1}^m \omega_{(k,\cdots,n)} \Delta z_k \right)
\right]\; \right) \;, \quad \qquad  
\label{difdistro} 
\end{eqnarray}
where $\sum_2^1 \equiv 0$ is understood.  All vectors in 
Eq.~(\ref{difdistro}) are two dimensional and coincident 
with the plane orthogonal to the direction of jet propagation.
We denote by $\hat{\bf \delta}$ the unit vector in the direction of
the radiative gluon's transverse momentum, ${\bf k} = \omega 
\sin \theta^* \hat{\bf \delta}$.  
The color current propagators read
\begin{eqnarray}
{\bf C}_{(m, \cdots ,n)} &=&  \frac{1}{2} \nabla_{\omega 
\sin \theta^* \hat{\bf \delta} } 
\ln \, ( \omega \sin \theta^* \hat{\bf \delta}
 - {\bf q}_m - \cdots  - {\bf q}_n )^2 \;, \nonumber \\  
{\bf B}_{(m+1, \cdots ,n)(m, \cdots, n)} &=&  
{\bf C}_{(m+1, \cdots ,n)} - {\bf C}_{(m, \cdots ,n)} \;\;. 
\label{props}
\end{eqnarray}
The momentum transfers ${\bf q}_i$ are distributed according to 
a normalized elastic differential cross section
\begin{equation}
\sigma_{el}(i)^{-1}\frac{d \sigma_{el}(i)}{d^2 {\bf q}_i}  
= \frac{\mu^2(i)}{\pi({\bf q}_i^2+\mu^2(i))^2} \; ,
\label{GWmodel}
\end{equation}
which models scattering  by  soft partons with a thermally 
generated Debye screening mass $\mu(i)$. 
From Eq.~(\ref{GWmodel}) 
$\langle {\bf q}_i^2 \rangle \propto \mu^2(z_i)$ and for a quark-gluon
plasma in local thermal equilibrium  $\mu^2(z_i) \sim 4 \pi \alpha_s 
T^2(z_i)$. It has been shown~\cite{Gyulassy:2000fs} via the 
cancellation of direct and virtual diagrams that in the 
eikonal limit only the gluon mean free path $\lambda_g(i)$ 
enters the medium-induced bremsstrahlung spectrum in  
Eq.~(\ref{difdistro}). For gluon dominated bulk soft matter 
$\sigma_{el}(i)  \approx  \frac{9}{2} \pi \alpha_s^2/\mu^2(i)$ 
and  $\lambda_g(i)=1/\sigma_{el}(i)\rho(i)$.  
For the case of (1+1)D dynamical Bjorken expansion of the 
QGP~\cite{Gyulassy:2000gk,Wang:2003mm}
\begin{equation}
 \mu(z_i) = \mu(z_0) \left(\frac{z_0}{z_i} \right)^{1/3} \!\!, 
\, \qquad \lambda_g(z_i) = \lambda_g(z_0) 
\left(\frac{z_i}{z_0} \right)^{1/3}\;.
\end{equation}

The characteristic path length dependence, $\propto L^2$  
for static plasmas,  of the non-Abelian energy 
loss in Eq.~(\ref{difdistro}) comes from the
interference phases and is differentially controlled
by the inverse formation times,    
\begin{equation}
\omega_{(m,\cdots,n)}  = 
\frac{( \omega 
\sin \theta^* \hat{\bf \delta} 
- {\bf q}_m - \cdots  - {\bf q}_n )^2}{ 2 \omega }  \;, 
\label{ftimes}
\end{equation}
and the separations of the subsequent scattering centers  
$ \Delta z_k = z_k - z_{k-1} $. It is the non-Abelian 
analogue of the Landau-Pomeranchuk-Migdal destructive
interference effect in QED~\cite{Landau:um}. The extension
to the case of massive quarks can be carried out 
systematically to all orders in 
opacity~\cite{Djordjevic:2003zk} by identifying the 
heavy quark mass terms $\propto \omega^2 M_q^2 / E^2 $.

It can be shown from  Eq.~(\ref{difdistro}) that the 
propagator poles, Eq.~(\ref{props}), where potential 
divergences may arise are precisely canceled by the 
interference phases, Eq.~(\ref{ftimes}), to all orders in 
opacity. This  
point will be better illustrated if we examine  the 
Gyulassy-Levai-Vitev~\cite{Gyulassy:2000fs} (GLV) gluon
distribution in angle and frequency  to first order in the 
mean number of soft interactions in the plasma. The diagrams
shown in Fig.~\ref{diags} yield:
\begin{eqnarray} 
|{\cal M}_a|^2 + 2 {\rm Re} {\cal M}_b^\dagger {\cal M}_c 
 & \propto & 
 \frac{dN^{g\,(1)}_{\rm med}}{d\omega d \sin \theta^*  d \delta } =   
\frac{2 C_R \alpha_s}{  \pi^2 } 
\int_{z_0}^L \frac{d \Delta z_1}{\lambda_g(z_1)}  
\int d^2 {\bf q}_1 \, \frac{1}{\sigma_{el}}  
\frac{d \sigma_{el}}{d^2 {\bf q}_1} (z_1)  
   \nonumber \\[.5ex] && \!\!\!\!\!  \times 
 \frac{  {\bf q}_1 \cdot \hat{\bf \delta}  }  { {\bf q}_1^2 - 
2  \omega \sin \theta^*  {\bf q}_1 \cdot \hat{\bf \delta} 
+ \omega^2 \sin^2 \theta^*} 
 \nonumber \\[1.ex] && \!\!\!\! \times 
\left[ 1 -  \cos \left( \frac{ ({\bf q}_1^2 - 
2  \omega \sin \theta^*  {\bf q}_1 \cdot \hat{\bf \delta} 
+ \omega^2 \sin^2 \theta^*) 
\Delta z}{2 \omega}  \right)  \right]  \;. \qquad  
\label{unintspect} 
\end{eqnarray}
In Eq.~(\ref{unintspect}) the two terms, $|{\cal M}_a|^2$
and  $2 {\rm Re} {\cal M}_b^\dagger {\cal M}_c$ are individually 
divergent. Their sum is, however, well defined due to the cancellation
of the soft and collinear singularities. 
From Eq.~(\ref{unintspect})  the gluon distribution 
is not only finite when $\theta^* \rightarrow 0$  but vanishes 
on average  due to the uniform angular distribution of momentum 
transfers from the medium, 
$\int_0^{2\pi} d \alpha \, \cos \alpha = 0$. Here   $\cos \alpha
= {\bf q}_1 \cdot \hat{\bf \delta} / | {\bf q}_1 |$.  
We have checked that for physical gluons of $|{\bf k}| \leq \omega$ 
the cancellations discussed  here persist to all orders in the 
mean number of  scatterings~\cite{Gyulassy:2000fs}.
The small frequency and small angle spectral behavior  
of $\, {dN^g_{\rm med}}/{d\omega d \sin \theta^*  d \delta } \,$ remains 
under perturbative control.

We also emphasize that destructive 
quantum interference suppresses radiation 
of $\Delta z \ll l_f$~\cite{Gyulassy:2000fs}. Gluons of final
transverse momentum ${\bf k}$ are initially much more collinear with
the jet but acquire their moderately large angle $\theta^*$ 
from momentum transfers from the medium.  
The formation length at the emission vertex, which allows for 
interference on the scale of the nuclear size,  reads  
\begin{equation}
 l_f = \frac{2 \omega}{(\bf{k} -\bf{q}_1 )^2}  = 
 \frac{2\omega}
{ {\bf q}_1^2 - 
2  \omega \sin \theta^*  {\bf q}_1 \cdot \hat{\bf \delta} 
+ \omega^2 \sin^2 \theta^* } \;.
\end{equation}
The induced radiation decouples from the jet 
at  $l_f \sim \Delta z  \propto L/2$ and reduces 
the possibility of gluon showering due to further interactions
in the plasma. For example, in central $L = 9$~fm collisions
$\langle \, l_f \, \rangle \simeq 4$~fm and with formation
times $ z_0 = 0.25 - 0.5$~fm by the time a physical gluon 
is formed the plasma density has dropped by a factor of 
8 to 16 from the longitudinal Bjorken expansion.

Given the vastly different angular behavior of the vacuum and 
the medium-induced gluon bremsstrahlung, Eqs.~(\ref{divsmall}) 
and (\ref{unintspect}), it is critical to identify
the phase space where cancellation of the color currents 
induced by the hard and soft scattering occurs. We fix the parameters
of the medium in Eq.~(\ref{unintspect}) 
to $\mu(z_0) = 1.5$~GeV and $\lambda_g(z_0) = 0.75$~fm
at initial formation time $z_0 = 0.25$~fm. Since small $|{\bf k}|$
and $\omega$  
emission is suppressed, we use only a moderate $\alpha_s = 0.25$. 
Triggering on high $p_{T_1}$ hadron  directs its parent parton ``c'' 
away from the medium and  places the collision point of the lowest order 
(LO) ${\rm ab} \rightarrow {\rm cd}$ underlying perturbative 
process~\cite{Owens:1986mp} 
close to the periphery of the nuclear overlap region.
Then, it is the back-scattered jet ``d'' that 
traverses the QGP.  For large nuclei, such as Au and Pb,  path lengths  
$L = 9 \; (3)$~fm are used to illustrate central (peripheral) 
collisions, respectively.  We limit gluon emission to the forward 
jet hemisphere,  $0 \leq \theta^* \leq \frac{\pi}{2}$.

The angular distribution of medium-induced radiation for 
$E = 6$~GeV gluon jet for select values of $\omega$ is shown in 
the right panels of Fig.~\ref{hole}. We find that gluon 
emission is strongly suppressed 
within a cone of opening angle $\theta^* \simeq 0.25$~rad due to 
the cancellation of collinear bremsstrahlung -- a mechanism
different from a Gaussian random walk in $\theta^*$.    
The broad gluon distribution can be characterized by the mean 
emission angle
\begin{equation}
 \langle  \theta^*  \rangle = {\int_0^1 
\theta^* \frac{dN^g_{\rm med_{}}}{d\omega d \sin \theta^* } 
\,d\sin \theta^*  }      \left[ {\int_0^1 
\frac{dN^g_{\rm med_{}}}{d\omega d \sin \theta^*_{} }_{} \, 
d\sin \theta^* }   \right]^{-1} \!\!\! \;, 
\end{equation}
given in the inserts of Fig.~\ref{hole}. Left panels illustrate
the infrared safety property of the spectrum. Energy loss is 
dominated by semihard few GeV gluons, see $\omega dN^g/d\omega$.
The gluon number $ dN^g / d \omega $ distribution peaks at even 
smaller frequencies $\omega$.  Momentum transfers from the 
medium $|{\bf q}_1| \sim 1 - 2$~GeV are clearly  able to generate     
broad angular bremsstrahlung distribution. However, it is the 
cancellation of the $\theta^* \rightarrow 0$ radiation together
with the model of the medium  which
ensure the weak $\omega$ dependence of the mean emission angle
$ \langle  \theta^*  \rangle \sim 0.5$~ rad seen in Fig.~\ref{hole}.

\section{Suppression of the single inclusive hadrons}

Dynamical nuclear effects in multi-particle production in A+B 
collisions can be studied through the ratio
\begin{equation}
\!  R^{(n)}_{AB}  =  \frac{d\sigma^{h_1 \cdots h_n}_{AB} / 
dy_1 \cdots dy_n d^2p_{T_1} \cdots d^2p_{T_n}} 
{\langle N^{\rm coll}_{AB} \rangle\, d\sigma^{h_1 \cdots h_n}_{NN} / 
dy_1 \cdots dy_n d^2p_{T_1} \cdots d^2p_{T_n}} \; .
\label{multi-q}
\end{equation}
In Eq.~(\ref{multi-q}) $\langle N^{\rm coll}_{AB} \rangle$ is
the mean number of binary collisions which can be calculated
from an optical Glauber model.

\begin{figure}[t!]
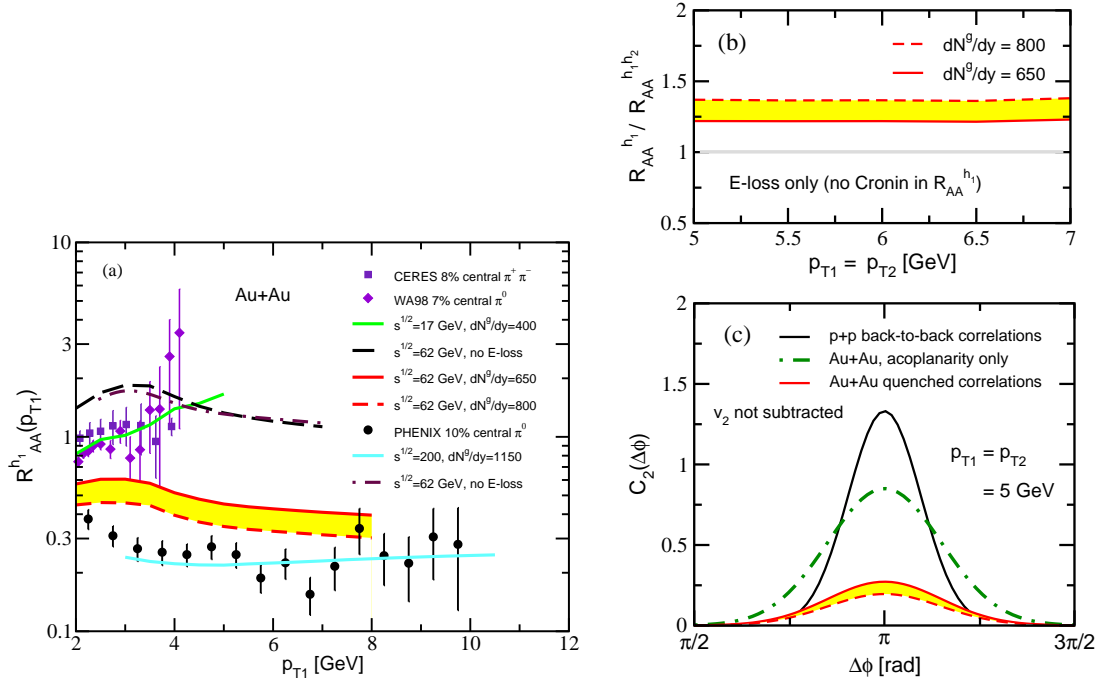

\begin{center} 
\includegraphics[width=3.in,angle=0]{Fig2.NEWa.eps}
\hspace{.1in}
\includegraphics[width=2.5in,angle=0]{Fig2.NEWb.eps}
\vspace*{0.in}
\caption{(a) Calculated nuclear modification factor  
$R_{AA}( p_{T_1})$ versus the  center of mass energy 
$\sqrt{s_{NN}}=17 , \, 62, \, 200$~GeV for central Au+Au collisions.
The yellow band represents the perturbative QCD expectation for the 
depletion of the neutral pion multiplicity
at $\sqrt{s_{NN}}=62$~GeV. 
(b) The ratio of the attenuation of the single and 
double inclusive $\pi^0$ cross sections 
$ R^{h_1}_{AA}(p_{T_1}) / R^{h_1h_2}_{AA}(p_{T_1}= p_{T_2})$ 
in central Au+Au collisions at $\sqrt{s_{NN}}=62$~GeV
and  (c) the  manifestation of the double inclusive hadron suppression 
in the quenching of the away-side di-hadron correlation function 
$C_2(\Delta \phi)$~\cite{Qiu:2004da}. }
\label{fig2}
\end{center} 
\end{figure}

Results form the perturbative calculation of the nuclear 
modification to the neutral and charged pion production in central 
$Au+Au$ collisions are given in the left hand side of 
Figure~\ref{fig2}. At all energies there is  cancellation between 
the Cronin enhancement, which arises from the transverse momentum 
diffusion of fast partons in cold nuclear 
matter~\cite{Gyulassy:2000gk,Vitev:2003xu,Qiu:2003pm,Zhang:2001ce}, 
and the subsequent inelastic jet attenuation in the  final state. 
This interplay is most pronounced at the low Super Proton
Synchrotron (SPS) $\sqrt{s_{NN}}=17$~GeV  
where, in the absence of quenching, the corresponding enhancement 
could reach a factor of 3-4~\cite{Gyulassy:2000gk}. With final state 
energy loss taken into account, $R^{h_1}_{AA}(p_{T_1})$ is 
shown versus the 
WA98~\cite{Aggarwal:2001gn} and the CERES~\cite{Agakishiev:2000bt} 
Pb+Pb and Pb+Au data with p+p baseline defined as 
in~\cite{d'Enterria:2004ig}. Systematic comparison of the 
calculated nuclear modification to the existing experimental 
measurements can also be found in~\cite{d'Enterria:2004ig}. The 
agreement between data and theory is good at high $p_{T_1}$ but 
the quenching is over predicted at $p_{T_1} \sim\;$few GeV. 

In two particle correlation measurements the energy loss 
of the second parent parton at high $p_{T_1},\, p_{T_2}$ 
qualitatively leads to $ 1 \leq R^{h_1}_{AA} / R^{h_1h_2}_{AA} \leq 2$.  
Numerical estimates in Fig.~\ref{fig2}(b) from jet quenching alone 
indicate that the double inclusive 
hadron suppression is $25\% - 40\%$ larger than the single inclusive 
quenching in the $5 \leq p_{T_1} = p_{T_2} \leq 7 $~GeV range  at
$\sqrt{s_{NN}}=62$~GeV.  Double inclusive cross 
section quenching is manifest  in the attenuation of the away-side 
correlation function  $C_2(\Delta \phi) = (1/N_{\rm trig}) 
dN^{h_1h_2}/d\Delta \phi $~\cite{Qiu:2004da}.  
In Fig~\ref{fig2}(c) it leads to a $4-5$ fold attenuation of the area 
$A_{\rm far}$ relative to the p+p case. In contrast, elastic 
transverse momentum diffusion will only result in a 
broader $C_2(\Delta \phi)$.  We note that the experimentally 
measured suppression  value will be sensitive to the subtraction 
of the elliptic flow  $v_2$ component
which may remove part or all of the residual correlations. 

To study the details of non-Abelian energy loss in the QGP we extend
the calculation of the quenching of the away-side di-hadrons to
low values of $p_{T_2}$ in the next section.

\section{Redistribution of the lost energy}

Results reported in  Ref.~\cite{Vitev:2005yg} 
are important for the future LHC heavy 
ion program since they for the first time suggest a large, and therefore 
detectable, broadening of the jet cone in the nuclear environment.
Note that  $ \langle  \theta^*  \rangle$ implies a redistribution of
the energy flow even  within jets of large opening angle  
$R = \sqrt{ \Delta \eta^2  + \Delta \phi^2}  = 1$. If high statistics
measurements of jet quenching away from midrapidity become available
at RHIC II novel ``octupol twist'' moments of the parton absorption
pattern in the QGP may be accessible~\cite{Adil:2005qn}.
At present, however, a key question for perturbative QCD phenomenology 
is whether the medium induced gluon bremsstrahlung can significantly 
alter the di-hadron correlations measured at midrapidity at 
RHIC~\cite{Adler:2002tq,unknown:2005ph,Rak:2004gk,Filimonov:2004hs}. 
We naturally 
focus on  the away-side $| \Delta \varphi | \geq \frac{\pi}{2}$ 
case, where medium effects are the largest.  Nuclear modifications  
build upon the LO double inclusive hadron production cross 
section, which is calculable in the perturbative QCD factorization 
approach~\cite{Collins:1985ue}  if either of the hadrons is 
moderately hard ($p_{{T_1}}{\rm \, or \;}  p_{{T_2}}\geq$~few 
GeV)~\cite{Owens:1986mp,Qiu:2004da}    
\begin{eqnarray} 
 \frac{ d \sigma^{h_1 h_2}_{NN} }{ dy_1  dy_2 
 dp_{T_1} dp_{T_2} d\Delta \varphi} &=& K
\sum_{abcd}  \int_0^1 \frac{dz_1}{z_1} \, D_{h_1/c}(z_1) \, 
\left[ D_{h_2/d} (z_2) \delta (\Delta \varphi - \pi)  \right] \,  
\frac{\phi_{a/N}({x}_a)\phi_{b/N}({x}_b)}{{x}_a{x}_b\, {S}^2 } \, 
\nonumber \\[.5ex] 
&& \times  \; 
2 \pi \alpha_s^2 |\overline {M}_{ab\rightarrow cd}|^2 \; .
\label{double}
\end{eqnarray}
In Eq.~(\ref{double}) $K=2$ is a next-to-leading order $K$-factor, 
$x_{a,b}=p_{a,b}/p_{N_a,N_b}$ are the momentum fractions of the 
incoming partons and $z_{1,2} = p_{h_1,h_2}/p_{c,d}$ are the 
momentum fractions of the hadronic fragments. We use standard 
lowest order Gluck-Reya-Vogt PDFs~\cite{Gluck:1998xa} and  
Binnewies-Kniehl-Kramer FFs~\cite{Binnewies:1994ju}. Renormalization,
factorization and fragmentation scales are suppressed everywhere 
for clarity. The spin (polarization) and color averaged 
matrix elements $|\overline {M}_{ab\rightarrow cd}|^2$ are given
in~\cite{Owens:1986mp}.

We first recall the physical effects 
that alter the LO perturbative formula, Eq.~(\ref{double}). 
A modification that does not change the $\Delta \varphi$-integrated 
cross section is vacuum- and medium-induced  
acoplanarity\cite{Qiu:2003pm,Rak:2004gk}. 
The deviation of jets from being 
back-to-back in a plane perpendicular to the collision axis 
arises from the soft gluon radiation
and  transverse momentum diffusion in dense nuclear 
matter~\cite{Qiu:2003pm}. In the approximation of collinear 
fragmentation, the width of the away-side hadron-hadron
correlation function can be related to the accumulated di-jet 
transverse momentum squared in the $\varphi$-plane,   
$ \sin \sqrt{\frac{2}{\pi}} \sigma_{\rm Far} = 
\sqrt{ \frac{2}{\pi}  \langle k_T^2 \rangle_\varphi} / p_{\perp_d}$. 
Assuming a Gaussian form, 
\begin{equation}
f_{\rm vac. \; or \; med.} ( \Delta \varphi ) = 
{ \frac{1}{\sqrt{2\pi} \sigma_{\rm Far} } } 
\exp \left[ -\frac{(\Delta \varphi - \pi)^2}{2\sigma^2_{\rm Far} } \right] \; ,
\end{equation}
a good description of $|\Delta \varphi| \geq \frac{\pi}{2}$ 
correlations measured in elementary p+p collisions at 
$\sqrt{S} = 200$~GeV~\cite{Adams:2003im} requires a large 
$\langle k_{T\; {\rm vac}}^2 \rangle_\varphi =  5\; {\rm GeV}^2$  
for the di-jet pair with  away-side scattered quark (and a 2.25 
larger value for a scattered gluon). Additional broadening arises 
from the interactions of the jet in the QGP that ultimately
lead to the reported energy loss. Using the parameters of the 
medium for (1+1)D expansion we find
\begin{equation} 
\langle k_{T\,\rm hot}^2 \rangle = \int_{z_0}^{L} dz \; 
2\frac{\mu^2_D(z)}{\lambda_{q,g}(z)} =  
2 \frac{\mu^2_D(z_0)}{\lambda_{q,g}(z_0)} \ln \frac{L}{z_0} \;,
\label{acoplan}
\end{equation} 
although only half  is projected on the $\varphi$-plane, 
$\langle k_T^2 \rangle_\varphi = 
\langle k_{T \; {\rm vac}}^2 \rangle_\varphi  
+ \frac{1}{2} \langle k_{T\; \rm hot}^2 \rangle$.

\begin{figure}[t!]
\begin{center}
\includegraphics[width=4.5in,height=3.3in,angle=0]{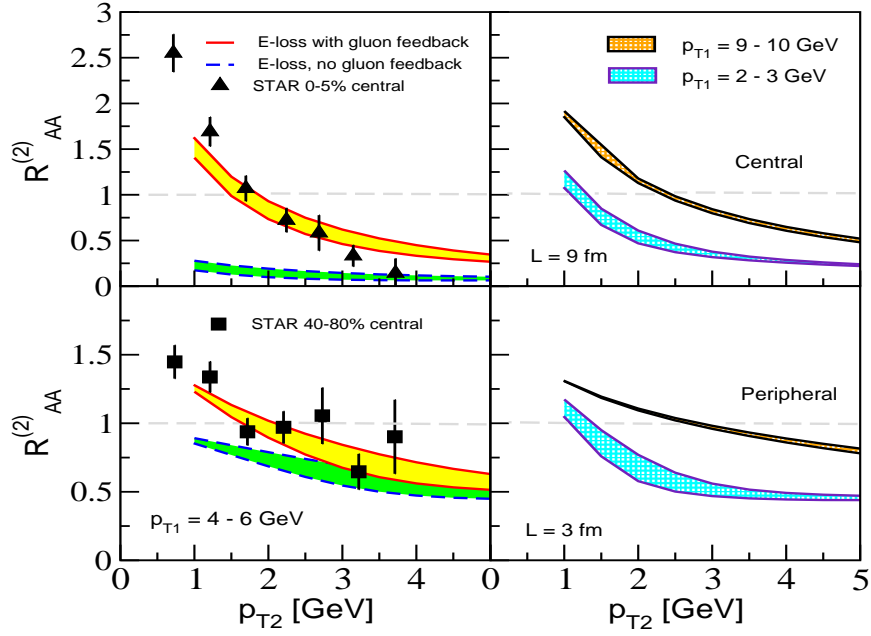}
\end{center}
\caption{  Nuclear modification of the back-to-back di-hadron 
correlations with and without the contribution of medium-induced 
bremsstrahlung. Top and bottom panels illustrate central and 
peripheral collisions, respectively. Data is from 
STAR~\cite{unknown:2005ph}. Right panels predict $R^{(2)}_{AA}$ for
low and high $p_{T_1}$ hadron triggers. }
\label{yields}
\end{figure}

Two competing mechanisms do, however, change the $p_{T_2}$ dependence 
of the perturbative cross section, Eq.~(\ref{double}). First is the 
the parent jet ``d'' fractional energy loss 
$\epsilon = \Delta E_d / E_d$,  which we here for 
simplicity consider on average and evaluate by integrating 
Eq.~(\ref{unintspect}). It leads to a rescaling of 
the hadronic fragmentation momentum fraction  
$z_2 \rightarrow z_2 / (1 - \epsilon )$~\cite{Gyulassy:2000gk}.  
Physically, fewer high $p_{T_2}$ particles are produced by 
the attenuated parton of energy $E_d - \Delta E_d$. 
If the energy loss is large, a second 
mechanism is invoked as a consequence. Hadronic fragments of the 
radiative gluons will increase the probability of finding  
low and  moderate  $p_{T_2}$ particles associated with the 
interacting jet~\cite{Pal:2003zf}.

To calculate di-hadron correlations, we first map the jet structure 
of a hard $90^0$-\,scattered parton on  rapidity  
$y \approx \eta = - \ln \tan (\theta /2)$  ($\theta$ being the angle 
relative to the  beam axis) and azimuth $\phi$,   
$ \tan^2 \theta^* = \cot^2  \theta + \tan^2 \phi \, , \; 
\tan \delta  = - { \cot \theta}/{\tan \phi } . $
The approximately flat rapidity distribution of the away-side jet 
near $y_2=0$ can be used to sum over all emission angles 
$\theta \in (\theta_{\rm min}, \theta_{\rm max}) \subset 
(0,\pi)$ yielding 
\begin{equation}
\frac{dN^g_{\rm med}}{d \omega d \phi} = 
\int\limits_{\theta_{\min}}^{\theta_{\max}}  d \theta  \;
\left[ \, \frac{dN^g_{\rm med}}{d \omega d \sin \theta^* d \delta }  
\left| \frac{\partial(\sin \theta^*, \delta ) }{\partial(\theta , \phi )} 
  \right| \; \right]  \; .
\label{trans}
\end{equation} 
The Jacobian of the transformation in Eq.~(\ref{trans}) reads    
\begin{equation}
    \left| \frac{\partial(\sin \theta^*, \delta ) }
    {\partial(\theta , \phi )}  \right|
    = \frac{1}{\sin^2 \theta  \cos^2 \phi}
    \frac{ (\tan^2 \phi + \cot^2  \theta )^{-1/2} } 
  {( 1 + \tan^2 \phi + \cot^2 \theta )^{3/2} }     \;. 
\end{equation} 
It is critical to note that projection on a plane 
coincident with the jet cone axis  (the $\phi$-plane in 
Eq.~(\ref{trans}) is one example) efficiently masks the  
$\theta^* \rightarrow 0$ 
void in the angular distribution of medium-induced gluons 
reported in Fig.~\ref{hole}. While projecting the medium induced
radiation on the $y=0$ plane is an approximation with the 
largest deviations being close to the edges of phase space 
(in this case  $\frac{\pi}{2}$ and  $\frac{3\pi}{2}$ ),
it is much less expensive computationally than a full simulation 
of the away-side jet+gluons.  
Our conclusion about the importance of the broad rapidity 
distribution of di-jets and the vacuum and medium-induced
acoplanarity is 
independent of the physical mechanism that depletes the 
parton (or particle) multiplicity in a cone around the 
jet axis.

The end analytic result for the modification 
to Eq.~(\ref{double}) per average nucleon-nucleon collision 
in the heavy ion environment  can be derived  from the energy 
sum rule for all hadronic fragments from the jet,
\begin{eqnarray}
D_{h_2/d} (z_2) \delta (\Delta \varphi - \pi) 
 & \Rightarrow & 
\frac{1}{1-\epsilon} D_{h_2/d}\left( \frac{z_2}{1-\epsilon} \right) 
f_{\rm med.}( \Delta \varphi ) 
+  \, \frac{p_{T_1}}{z_1}    \int_0^1 \frac{d z_g}{z_g}  
D_{h_2/g}(z_g) 
\nonumber \\  &&   \times \int_{-\pi/2}^{\pi/2} d \phi  \;
\frac{ dN^g_{\rm med} (\phi) }{ d\omega  d  \phi }  
f_{\rm vac.}( \Delta \varphi - \phi) \;. 
\label{nucmod}
\end{eqnarray}
Here, $z_g = p_{T_2} / \omega $.

Whether medium-induced gluon radiation may have significant observable
consequences for the large angle di-hadron correlations depends on its
relative contribution to the $\Delta \varphi$-integrated cross section. 
From Eq.~(\ref{double}) this can be studied versus $p_{T_2}$ via the 
ratio $ R^{(2)}_{AA}$~\cite{Qiu:2004da}.  
Numerical results, 
shown in Fig.~\ref{yields}, correspond to triggering on 
a high $p_{T_1} =\;$4 - 6~GeV pion and measuring all associated
 $\pi^+ + \pi^0 + \pi^-$.  Depletion of hadrons from  
the quenched parent parton alone leads to a large suppression of the 
double inclusive cross section with weak 
$p_{T_2}$ dependence.  Hadronic feedback from the  
medium-induced gluon radiation, however, completely changes 
the nuclear modification factor $R^{(2)}_{AA}$.     
It now shows a clear transition from a quenching of 
the away-side jet at high transverse momenta to enhancement
at  $p_{T_2} \leq  2$~GeV, a scale significantly 
larger than the one found in~\cite{Pal:2003zf}.  
In fact, we have checked that for 
large $\Delta E_d$ the back-to-back di-hadron correlations 
are dominated by radiative gluons to unexpectedly high
$p_{T_2} \sim 10$~GeV. It should be noted that the experimental
trigger introduces a bias toward large $z_1$, which is difficult
to account for analytically. Therefore we expect deviations
between theory and the measurement when $p_{T_2} \geq  p_{T_1}$. 
STAR data in central 
and peripheral Au+Au collisions~\cite{unknown:2005ph} is 
shown for comparison. Experimentally testable predictions 
for the shape and magnitude of $R^{(2)}_{AA}$ and 
the enhancement-to-suppression transition 
$R^{(2)}_{AA} = 1$ versus $p_{T_1}$ are also given in  
Fig.~\ref{yields}. 
It has been previously 
argued~\cite{Wang:2001cs} that detailed balance 
reduces jet suppression at partonic scales 
$p_{\perp_d} \leq \, {\rm few} \, \mu_D$, though such mechanism 
cannot produce enhancement of the low transverse momentum 
particle production observed in Fig.~\ref{yields}.

\begin{figure}[!t]
\begin{center}
\includegraphics[width=4.5in,height=3.3in,angle=0]{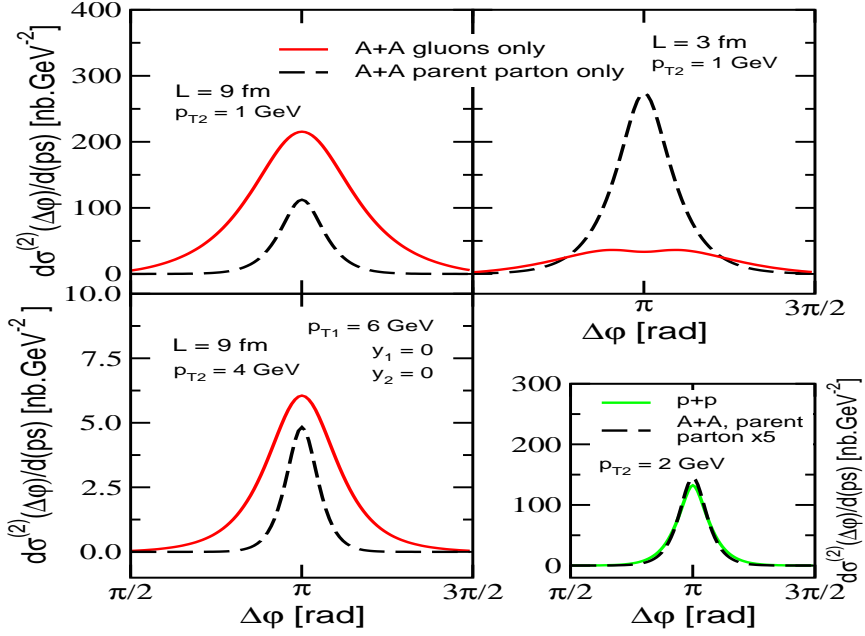}
\end{center}
\caption{ The angular  distribution of   
$|\Delta \varphi|\geq \frac{\pi}{2}$ di-hadrons for central and 
peripheral A+A collisions. Solid and dashed lines give {\em separately} 
the contribution of the radiative gluons and the attenuated parent 
parton. The lower right panel compares the the width of the correlation
function in p+p to the one of the suppressed parent parton in 
central A+A  reactions.}
\label{corelations}
\end{figure}

With a dominant contribution to the di-hadron yields, the 
medium induced gluons are bound to also determine the width of 
the correlation function. Two-particle distributions in A+A reactions,  
calculated  from Eqs.~(\ref{double}) and (\ref{nucmod}) 
for a $p_{T_1} = 6$~GeV trigger pion  and two different 
$p_{T_2}=1,\;4$~GeV associated pions,  are shown in 
Fig.~(\ref{corelations}).  Qualitatively, the medium-induced 
gluon component to the cross section controls the growth 
of the correlation width in  central and semi-central nuclear 
collisions. Quantitatively, the effect should be even larger than    
the one estimated here, which is limited by the 
imposed  $ 0 < \theta^* < \frac{\pi}{2}$ constraint. 
The decisive role of the bremsstrahlung spectrum, 
Eq.~(\ref{unintspect}), in establishing the $\Delta \varphi$-shape 
of away-side di-hadrons is further clarified in the bottom
right panel of Fig.~\ref{corelations}. Hadronic fragments from 
the quenched parton ``d'' {\em only} are shown to yield  a 
distribution that is {\em not} broader than the one anticipated  
in p+p reactions. Experimental measurements of significantly 
enhanced widths for $| \Delta \varphi| \geq \frac{\pi}{2}$  
two-particle correlation in A+A collisions should thus point 
to copious hadron production from medium-induced 
large angle gluon emission.

\section{Conclusions}

In summary, we calculated the transverse momentum and angular 
distribution of the away-side di-hadron correlations in the 
framework of the perturbative QCD factorization approach, 
augmented by inelastic jet interactions in the quark-gluon plasma. 
At RHIC energies we found that the medium-induced gluon radiation  
determines the two-particle yields and the width of 
their correlation function to surprisingly high transverse 
momentum $p_{T_2} \sim 10$~GeV. Clear transition from back-to-back 
jet enhancement to back-to-back jet quenching is established at 
moderate $ p_{T_2}  =  1 - 3 $~GeV, independent of collision 
centrality but sensitive to the trigger hadron momentum $p_{T_1}$.
Definitive experimental determination of its features,
predicted in Ref.~\cite{Vitev:2005yg}, will for the first time provide 
a handle on the {\em differential spectrum} of medium-induced
non-Abelian bremsstrahlung. 
Coincidental confirmation of large broadening of the away-side 
di-hadrons would require a critical reassessment of the origin 
of intermediate transverse momentum particles in central and 
semi-central nuclear collisions. A full numerical simulation
of jets away from midrapidity is required for more quantitative 
studies of the two particle topology. For jet physics at 
the LHC, our findings ensure a measurable increase in the jet 
width in ultradense nuclear matter.

\ack 
This work is supported by the J. Robert  Oppenheimer Fellowship
of the Los Alamos National Laboratory and by the US Department of Energy.

\end{document}